\documentclass[twocolumn,prb,showpacs,superscriptaddress,showkeys,preprintnumbers,amsmath,amssymb]{revtex4-2}

\usepackage{graphicx}
\usepackage{dcolumn}
\usepackage{bm}
\usepackage{multirow}
\usepackage{nicefrac}
\usepackage{braket}
\usepackage{amsmath, amssymb}
\usepackage{verbatim}
\usepackage{xcolor}
\usepackage[colorlinks=true,linkcolor=blue, citecolor=blue, urlcolor=blue,
unicode=true]{hyperref}
\usepackage{chemformula}
\usepackage{physics}
\usepackage[capitalise]{cleveref}
\graphicspath{{figures/}} 

\newcommand\ut[1]{_\mathrm{#1}}
\newcommand\tnone{\ensuremath{T\ut{N1}}}
\newcommand\tntwo{\ensuremath{T\ut{N2}}}

\begin{document}


\title{Possible stripe phases in the multiple magnetization plateaus in \ch{TbB4} derived from single-crystal neutron diffraction under pulsed high magnetic fields}

\author{N. Qureshi}

\email[Corresponding author:~]{qureshi@ill.fr}
\affiliation{Institut Laue-Langevin, 71 avenue des Martyrs, CS 20156,  38042 Grenoble Cedex 9, France}

\author{F. Bourdarot}
\affiliation{Service de Mod\'elisation et d'Exploration des Mat\'eriaux,
	Universit\'e Grenoble Alpes et Commissariat \`a l'Energie Atomique, INAC, 38054 Grenoble, France}

\author{E. Ressouche}
\affiliation{Service de Mod\'elisation et d'Exploration des Mat\'eriaux,
	Universit\'e Grenoble Alpes et Commissariat \`a l'Energie Atomique, INAC, 38054 Grenoble, France}

\author{W. Knafo}
\affiliation{CNRS, Laboratoire National des Champs Magn\'etiques Intenses,
	Univ. Grenoble Alpes, Univ. Toulouse 3, INSA Toulouse, EMFL, 31400 Toulouse, France}

\author{F. Iga}
\affiliation{Institute of Quantum Beam Science, Ibaraki University, Mito 310-8512, Japan}

\author{S. Michimura}
\affiliation{Research and Development Bureau, Saitama University, Saitama 338-8570, Japan}
\affiliation{Graduate School of Science and Engineering, Saitama University, Saitama 338-8570, Japan}

\author{L.-P. Regnault}
\affiliation{Institut Laue-Langevin, 71 avenue des Martyrs, CS 20156,  38042 Grenoble Cedex 9, France}

\author{F. Duc}
\email[Corresponding author:~]{fabienne.duc@lncmi.cnrs.fr}
\affiliation{CNRS, Laboratoire National des Champs Magn\'etiques Intenses,
	Univ. Grenoble Alpes, Univ. Toulouse 3, INSA Toulouse, EMFL, 31400 Toulouse, France}

\date{\today}

\begin{abstract}

We present a single-crystal neutron diffraction study on the Shastry-Sutherland lattice system TbB$_4$ at zero magnetic field and under pulsed high magnetic fields up to 35 T applied along the crystallographic $c$ axis. While our results confirm the magnetic structures at zero-field as well as those at the half- and full-magnetization plateaus, they offer new insight into the $\nicefrac{2}{9}$- and $\nicefrac{1}{3}$-magnetization plateaus observed in this system.
A stripe model of polarized 4-spin-plaquettes whose stripe density proportionally increases with the macroscopic magnetization is in full agreement with the neutron diffraction data. Equally well suited alternative models exist which explain the observed Bragg peaks which are inherently limited in a pulsed high magnetic field experiment. We discuss the different intensity distribution in $Q$ space which can be used to distinguish these models in future experiments.

\end{abstract}


\maketitle

\section{Introduction}
\label{sec:Introduction}

\begin{figure}
	\includegraphics[width=0.45\textwidth]{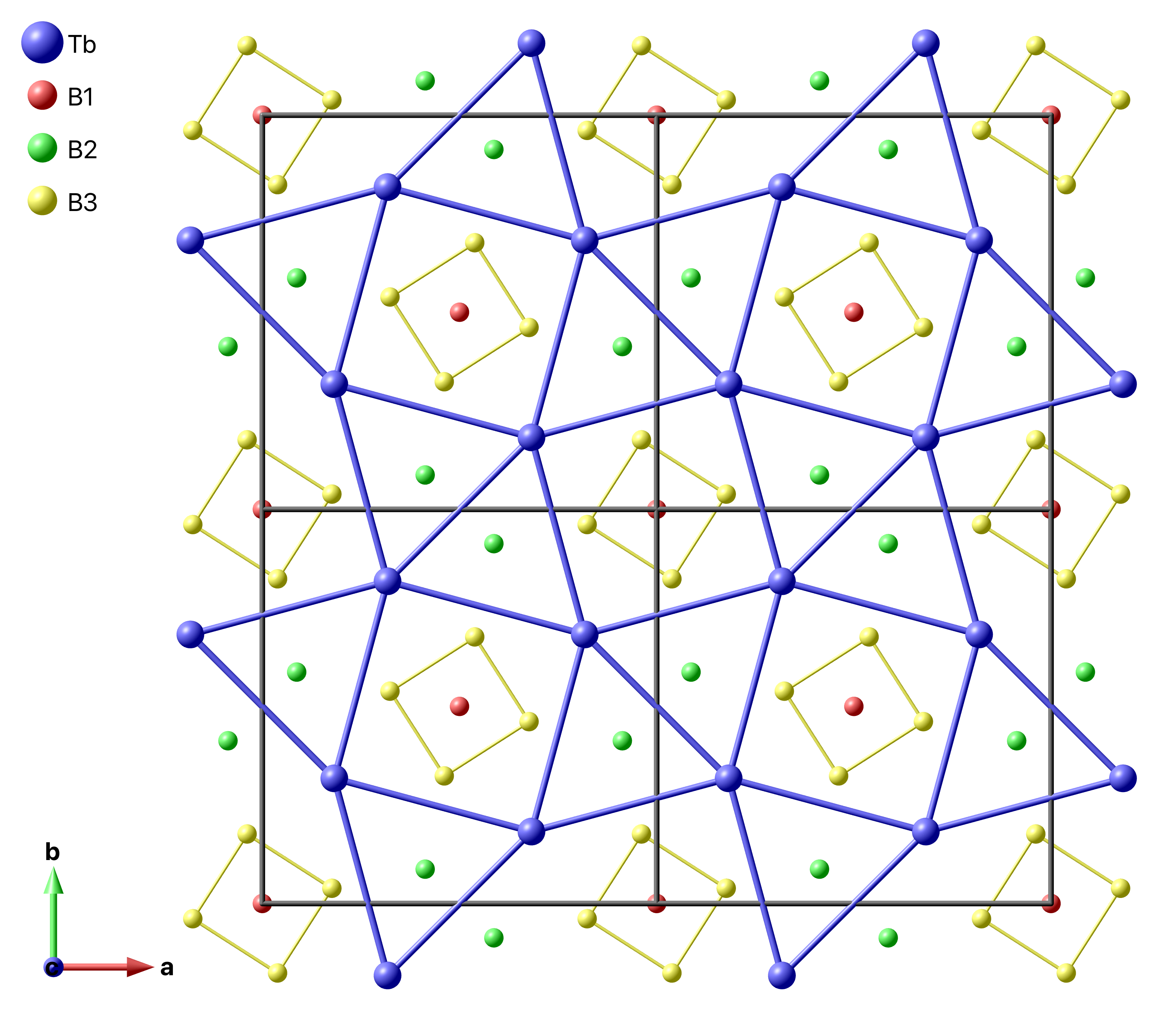}
	\caption{\label{fig:structure} View along the $c$ axis of the crystal structure of TbB$_4$ which maps onto the Shastry-Sutherland lattice consisting of connected triangles and squares (emphasized by blue bonds between the Tb ions). The cell edges are shown as black lines (a range from -0.2 to 2.2 expressed in multiples of lattice parameters is depicted along the $a$ and $b$ directions).}
\end{figure}

\begin{figure}
	\includegraphics[width=0.49\textwidth]{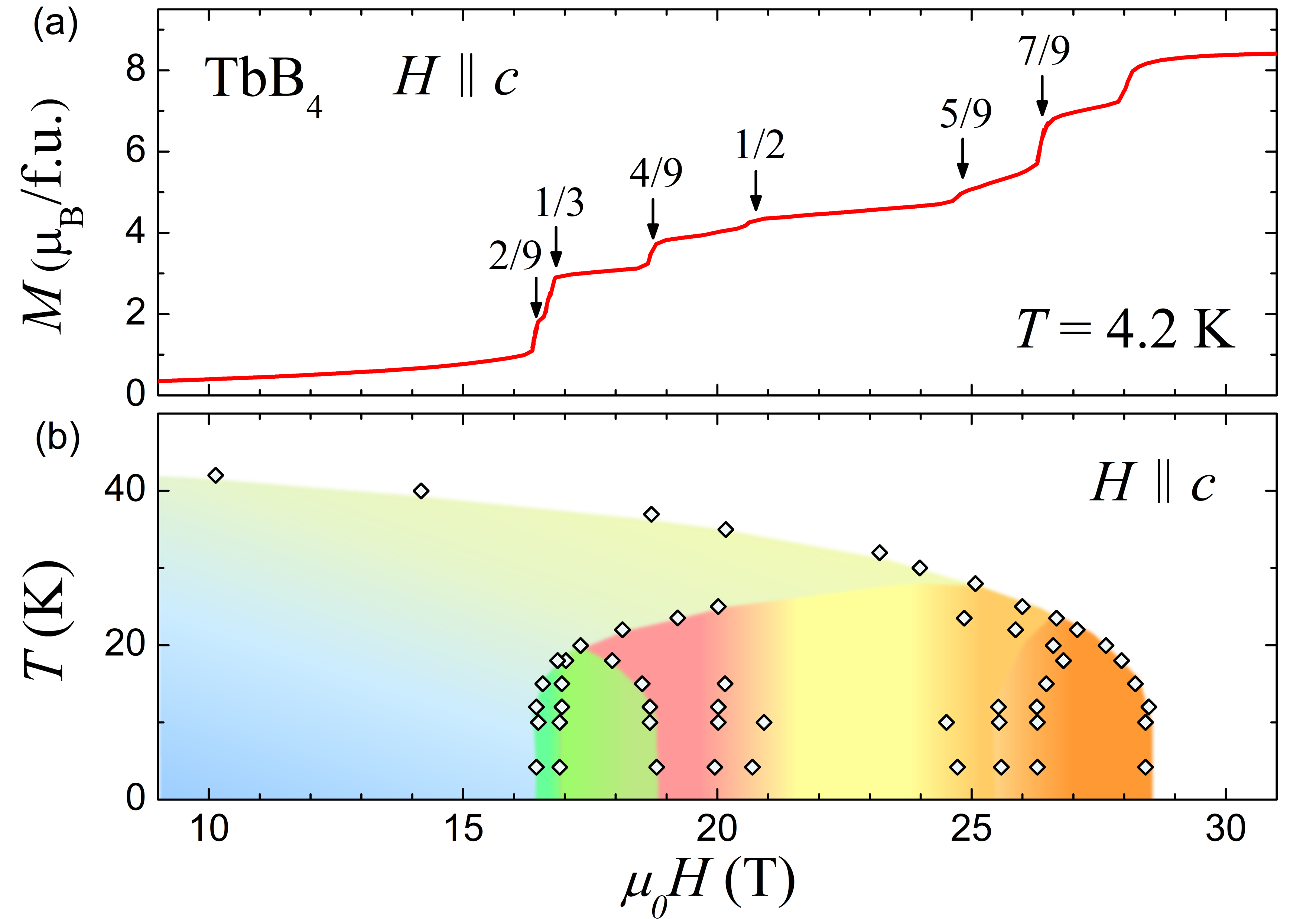}
	\caption{\label{fig:pd}  (a) Magnetization vs. magnetic field at $T$ = 4.2\,K for an applied magnetic field along the $c$ axis.
The arrows and fractions corresponding to the magnetization ratio $M/M_S$ indicate the transitions into the corresponding fractionalized magnetization plateaus. (b) Temperature vs. magnetic field phase diagram deduced from these magnetization measurements ((a) and (b) are both adapted from Ref.~\cite{yos2008}).}
\end{figure}

Geometrically frustrated systems with structural motifs such as triangles, squares or tetrahedra reveal strongly competing exchange interactions which lead to the suppression of long-range magnetic order, a large ground state degeneracy and exotic magnetic structures \cite{ram1994,moe2006}.
The Shastry-Sutherland lattice (SSL) is a well-known example of a frustrated system with an exact ground state solution and consists of a square lattice with antiferromagnetic nearest-neighbor  and alternating diagonal next-nearest-neighbor interactions \cite{sha1981}. The family of tetraborides crystallizing in a tetragonal space group $P4/mbm$ \cite{eto2006} has gathered a lot of interest as their crystal structure - consisting of a network of squares and triangles (see Fig.~\ref{fig:structure}) - can be mapped onto the SSL.
Diverse magnetic properties have been observed in the different members of the rare-earth ($R$) tetraborides $R$B$_4$ featuring simple antiferromagnetic structures with magnetic moments along the $c$ axis for ErB$_4$ \cite{sch1976, wil1981} and DyB$_4$ \cite{sch1976} or within the $a$-$b$ plane for GdB$_4$ \cite{fer2005,bla2006} and TbB$_4$ \cite{elf1981,wil1981,mat2007}, but also more complex magnetic structures like in HoB$_4$ \cite{oku2008} and TmB$_4$ \cite{sie2008,mic2009}. All these systems show an even more intricate behavior when a magnetic field is applied as evidenced by the presence of fractional magnetization plateaus \cite{yos2007,yos2008,bla2006,bru2018,ina2009,mat2010,bru2017} and substantial theoretical effort was made to identify the microscopic driving forces for these phenomena \cite{mol2009,hua2012,huo2013,gre2013,dub2013}. Neutron diffraction studies examining the plateau phases are rather scarce and focus on those rare-earth tetraborides with a low saturation field which is the case for the Ising-like HoB$_4$ \cite{bru2017} and TmB$_4$ \cite{sie2008}, with the exception of the first application of pulsed high magnetic fields to reveal the half-magnetization plateau in TbB$_4$ \cite{yos2009}, the compound being the focus of this study. \newline\newline
TbB$_4$ reveals two magnetic phase transitions at \tnone\ = 44 K and \tntwo\ = 24 K as shown by macroscopic methods \cite{fis1981,nov2013} and neutron diffraction experiments \cite{wil1981,mat2007}. An orthorhombic distortion to $Pbam$ symmetry was reported to take place at around 80 K \cite{hei1986}, i.e. above both magnetic phase transitions, but this was later challenged by Novikov \textit{et al.} \cite{nov2013} who located the structural transition temperature between \tnone\ and \tntwo. The magnetic structure below \tnone\ can be described in $P4/m'b'm'$ symmetry with the magnetic moments lying along the diagonals of the tetragonal basal plane and below \tntwo\ the spins tilt towards the $a$ axis within $Pb'a'm'$ symmetry \cite{mat2007} leading to two magnetic domains. Magnetization and magnetostriction measurements~\cite{yos2008} performed upon the application of a high magnetic field along the $c$ axis revealed a cascade of field-induced magnetic phase transitions and a complex phase diagram, which we reproduce in Fig.~\ref{fig:pd} due to its importance in the present study. The magnetic structure of the half-magnetization plateau $M/M_S=\nicefrac{1}{2}$ phase was determined by neutron diffraction experiments \cite{yos2009} in pulsed magnetic fields up to 30\,T. A model consisting of a $XY$- and Ising spin mixture was proposed in which the magnetic unit cell is doubled along the $a$ and $b$ axes and only one out of two 4-spin square-plaquette is significantly polarized along the $c$ axis. This model is motivated by the orthogonal spin arrangement within one plaquette (next-nearest neighbors) at zero field and the nearly orthogonal one along the diagonal interaction (nearest neighbours) with applied field suggesting the presence of a biquadratic term in the SSL Hamiltonian which stabilizes perpendicular magnetic moments and qualitatively explains the observed magnetization plateaus. \newline\newline
In this work, we present further single-crystal neutron diffraction data under a pulsed magnetic field applied along the $c$ axis yielding unique microscopic evidence for the $\nicefrac{2}{9}$- and $\nicefrac{1}{3}$-magnetization plateaus. The identified phases can be explained by a stripe model reminiscent of the charge order in cuprates \cite{tra1995,abb2005,wu2011,ghi2012,com2015} and manganites \cite{che1996,che1997,mor1998,che1999} as well as with the layered 214 nickelates \cite{tra1995b,sac1995,tra1996,woc1998} and cobaltates \cite{lan2014,wil2016,bab2016}. In the field-induced phases of TbB$_4$ the building block of a stripe consists of a polarized 4-spin plaquette belonging to one conventional crystallographic unit cell and the stripe density is proportional to the macroscopic magnetization value. We further discuss different potential models to explain both our and previously published data \cite{yos2009}. We detail how further neutron diffraction experiments are needed to distinguish the pertinent model.

\section{Experimental}
\label{sec:experimental}

Single crystals of TbB$_4$ enriched up to 99.52 \% by $^{11}$B were grown by the floating zone method detailed in \cite{iga1998}.
The single-crystal diffraction experiment at zero magnetic field was performed at the D23 diffractometer (ILL, Grenoble) in four-circle geometry using a plaquette-like sample with dimensions of roughly 2 x 2 x 0.5 mm$^3$ along the main crystallographic axes (note that the exact sample shape was described as a convex-hull model with 7 delimiting crystal planes using the \textsc{Mag2Pol} \cite{mag2pol} program which was also employed for the nuclear and magnetic structure refinement). A wavelength of 1.272~\AA\ provided by the (200) reflection of a Cu monochromator was used. The non-negligible absorption due to the imperfect B substitution was taken into account by calculating the beam path lengths inside the crystal ($\tau_{in}$ and $\tau_{out}$ representing the path lengths before and after the diffraction process, respectively) for all measured reflections in order to apply the transmission factor integral $\exp[-\mu (\tau_{in}+\tau_{out})]$. Hereby, the linear absorption coefficient $\mu$ depends on the refinable $^{11}$B concentration and is recalculated in every iteration of the structure refinement using the \textit{on-the-fly} absorption correction in \textsc{Mag2Pol}.\newline
The magnetic structures developing under a magnetic field applied parallel to the crystallographic $c$ axis were investigated on the CEA-CRG thermal neutron spectrometer IN22 (ILL, Grenoble). The instrument was equipped with the pulsed field set-up described in Ref.~\cite{Duc2018}, including a 1.15\,MJ generator and a 40-T conical pulsed magnet. The latter yields long pulses with a total duration of 100 ms, a rise time of 23 ms and a repetition rate of one 40-T shot every 10 min. A wavelength of $\lambda = 1.53$~\AA\ supplied by the (002) reflection of a PG monochromator was used for the experiment. Two different single-crystal samples were prepared for the measurements of the \mbox{($h$00)} and the \mbox{(110)} reflections, respectively, both specimens being of dimensions 2 x 2 x 1-2.5 mm$^3$.

\section{Results}
\label{sec:results}

\subsection{Zero-field magnetic structures}
\label{sec:zerofield}

The nuclear structure was investigated at \mbox{$T$ = 54 K} using a dataset consisting of 1565 Bragg reflections (146 unique in space group $P4/mbm$). As a first step the observed integrated intensities were averaged over $P\bar{1}$ symmetry (907 unique reflections) as Friedel pairs reveal exactly the same neutron beam path lengths. The refined parameters were the atomic positions, the isotropic temperature factors, the diagonal elements of the extinction correction tensor within an empirical \textsc{Shelx}-like model \cite{shelx}, an overall scale factor and most importantly the $^{11}$B occupation with that of the natural one being constrained to 1-$^{11}$B$_{occ}$. A convincing agreement ($R\ut{F}=6.06$) was obtained with a $^{11}$B occupation of 0.981(1). By setting the B occupation to the refined value the original dataset was corrected for absorption and averaged in $P4/mbm$ symmetry (146 unique reflections). The structure refinement yields a very good agreement factor of $R\ut{F}=4.28$ and the resulting parameters are shown in Tab.~\ref{tab:nucparams}. The self-consistency of the absorption correction was verified by refining the $^{11}$B occupation against the corrected dataset which yields 0.968(8) and only a marginal improvement of $R\ut{F}=4.26$.\newline

\begin{table}
	\caption{Refined structural parameters of TbB$_{4}$ within space group $P4/mbm$. Tb occupies Wyckoff position 4g \mbox{($x$ $x+\nicefrac{1}{2}$ 0)}, whereas the B ions are situated on positions 4e \mbox{(0 0 $z$)}, 4g and 8j \mbox{($x$ $y$ $\nicefrac{1}{2}$)} (in the same order as shown in the table). The $^{11}$B concentration was refined to 0.981(1) with the remainder being natural B. Note that the isotropic temperature factor was constrained to be equal for all B sites.}
	\label{tab:nucparams}
	\begin{ruledtabular}
		\begin{tabular}{ccccc}
			Atoms   &  $x$   &  $y$  &  $z$  &  $B$ (\AA$^2$)  \\ \hline
			Tb    &  0.3175(2)  &  0.8175(2)  &  0  &  0.19(4)  \\
			B1    &  0  &  0  &  0.2026(4)  &  0.37(4)  \\
			B2    &  0.0871(2)  &  0.5871(2)  &  $\nicefrac{1}{2}$  &  0.37(4)  \\
			B3    &  0.1763(2)  &  0.0386(2)  &  $\nicefrac{1}{2}$  &  0.37(4)  \\  \hline
			\multicolumn{5}{c}{Extinction parameters} \\
			\multicolumn{5}{c}{$x_{11}$ = 0.16(2)\quad $x_{22}$ = 0.19(3) \quad $x_{33}$ = 0.60(3)} \\
		\end{tabular}
	\end{ruledtabular}
\end{table}

\begin{figure}
	\includegraphics[width=0.35\textwidth]{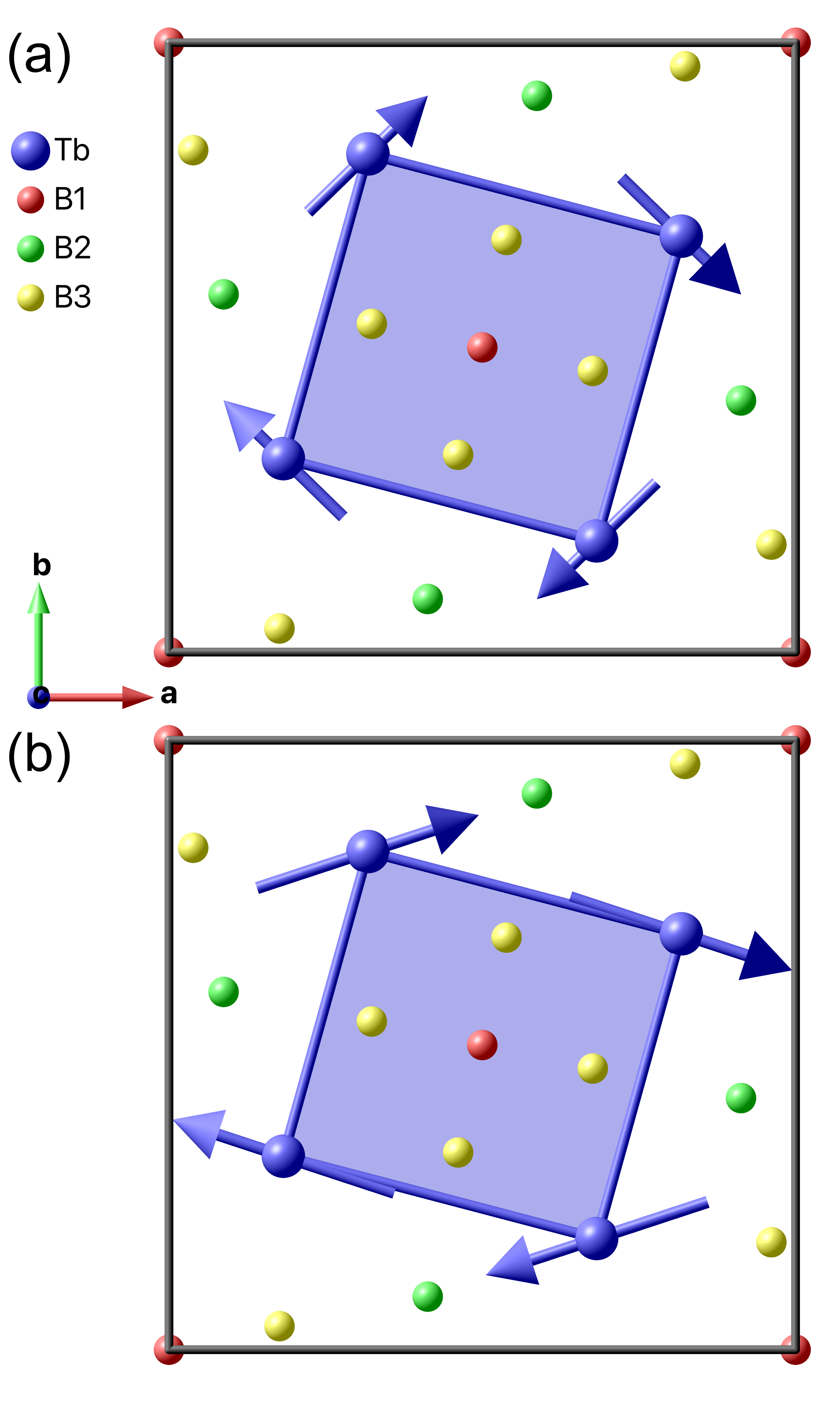}
	\caption{\label{fig:magstruct} Magnetic structures of TbB$_4$. (a) At 35 K the magnetic moments are aligned along the diagonal of the $a$-$b$ plane, while (b) at 11 K they tilt towards the $a$ axis by about 27$^\circ$.}
\end{figure}

The magnetic structure between \tnone\ and \tntwo\ was derived at $T$ = 34 K by analyzing 1564 Bragg reflections which were averaged in $Pmmm$ symmetry taking into account the reported nuclear structure (space group $Pbam$) and magnetic structure (broken $a$ and $b$ glide planes) yielding 302 unique reflections. In absence of a tabulated magnetic form factor for the Tb$^{4+}$ ion the analytical approximation to the $\langle j_0 \rangle$ integrals for the $f$ electrons of the Tb$^{3+}$ ion was used to describe the magnetic form factor throughout the data analysis of this work. By converting the obtained $P4/mbm$ structure to $Pbam$, fixing all nuclear structure, extinction as well as scale parameters to those obtained at $T$ = 54 K and by only refining two in-plane coefficients (multiplication coefficients to the basis vectors of the irreducible representation) of the magnetic moments in $Pb'a'm'$ symmetry an agreement factor of $R\ut{F}=4.72$ is obtained. The refined components are $C_1 = 4.59(5)~\mu\ut{B}$ along the $a$ axis and $C_2 = 4.47(4)~\mu\ut{B}$ along the $b$ axis, respectively, yielding a total magnetic moment of 6.41(9)$~\mu\ut{B}$ at 0.7(4)$^\circ$ from the [110] direction in full agreement with the reported magnetic structure (note that constraining the magnetic moment to lie exactly on the diagonal results in an amplitude of 6.39(6)~$\mu\ut{B}$ with no considerable improvement of the agreement factor). The resulting magnetic structure is shown in Fig.~\ref{fig:magstruct}(a).\newline
The ground state magnetic structure was studied at 11 K, i.e. below \tntwo. The same symmetry model was employed as for the magnetic structure at $T$ = 34 K including the same fit parameters. In a first step, we observe a significant tilt away from the diagonal with $C_1=6.44(7)~\mu\ut{B}$ and $C_2=5.68(6)~\mu\ut{B}$ ($R\ut{F}=6.06$) which led us to include a 90$^\circ$ twin in order to take into account those parts of the crystal in which the moments tilt in the opposite direction as a consequence of the tetragonal-to-orthorhombic phase transition. By doing so we obtain a clear improvement of the agreement factor ($R\ut{F}=5.33$) and an even more pronounced tilt which is expressed by $C_1=8.5(1)~\mu\ut{B}$ and $C_2=2.8(2)~\mu\ut{B}$ and amounts to roughly 27$^\circ$ which compares very well to the reported value in Ref.~\onlinecite{yos2009}. Refining the twin population only yields a marginal improvement ($R\ut{F}=5.22$) with a distribution of 0.55(9):0.45(9) for which an equally distributed twin population was fixed in the following. Note that introducing a 90$^\circ$ twin in the analysis of the magnetic structure in the \tntwo\ $<T<$ \tnone\ regime does not improve the refinement quality significantly which is in perfect agreement with a magnetic structure satisfying 4-fold rotation symmetry.

\subsection{Magnetization plateaus}
\label{sec:plateaus}

\begin{figure}
	\includegraphics[width=0.44\textwidth]{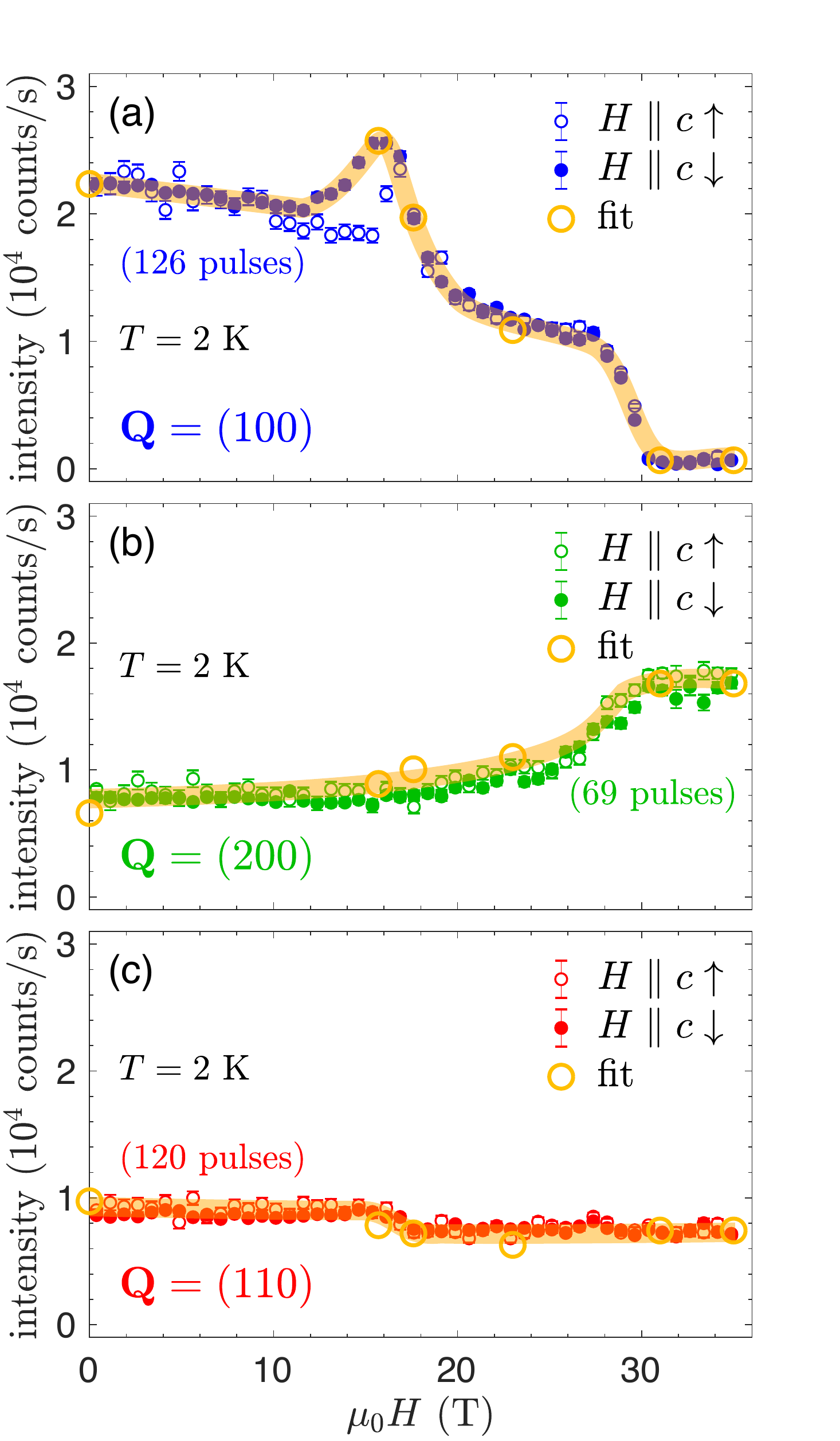}
	\caption{\label{fig:fielddep} Peak intensities of the (a) (100), (b) (200) and (c) (110) reflections as a function of increasing (open symbols) and decreasing (filled symbols) magnetic field applied along the $c$ axis. The number of accumulated pulses are mentioned in each graph. The orange circles denote the calculated peak intensities within the distinct phases which were investigated in detail (see main text) and the broad orange line is a guide to the eye joining the calculated points.}
\end{figure}

The magnetization plateaus were investigated by following the peak intensity  of selected Bragg reflections as a function of applied magnetic field.
The same reflections as in Ref.~\onlinecite{yos2009} were used, i.e. the purely magnetic (100) reflection as well as the (200) and (110) reflections
with both nuclear and magnetic contributions. Fig.~\ref{fig:fielddep} shows the field dependence of the neutron count accumulated and summed over a few tens
of pulsed-field shots and extracted with constant field-integration windows at $T$ = 2 K.
The bare comparison of the raw data with those in Ref.~\onlinecite{yos2009}, especially for the (100) reflection [Fig.~\ref{fig:fielddep}(a)], indicates an improvement in time resolution and a better control of the sample temperature. A sharp transition into the field-polarized state at $H\approx 28$ T and a clear anomaly around $H$ = 16 T corresponding to the $M/M_S=\nicefrac{2}{9}$ plateau are observed (note that the apparent hysteretic behavior is entirely related to the asymmetric sweep rate for increasing and decreasing fields \cite{Duc2018} and was not observed in the macroscopic measurements \cite{yos2008}). A monotonic decrease of the (100) peak intensity is observable up to 12 T, which can be explained by a small, but continuously increasing tilt of the magnetic moments towards the $c$ axis. Note that a ferromagnetic $c$ component respects the $Pbam$ symmetry and therefore does not contribute to the (100) reflection. Above 12 T the intensity steeply increases until it reaches its maximum at $H\approx$ 16 T, a value which fairly coincides with the first rise in magnetization up to $\nicefrac{2}{9}$ of the saturation value. The only way to describe this increase is to consider a larger $b$ component of the magnetic moments since the $a$ component is parallel to $\mathbf{Q}$ and therefore does not contribute to the magnetic scattering. Thus, independently of the details accounting for the field-induced magnetization, the magnetic moments are driven back towards the diagonal by a large enough field. A sharp drop of the (100) peak intensity marks the transition towards the $\nicefrac{1}{3}$-plateau which is followed by a further region of monotonically decreasing intensity corresponding to the $\nicefrac{1}{2}$-magnetization plateau.

Considering that only the in-plane spin components yield a non-zero contribution to the magnetic structure factor, the decrease of intensity by 50\% between the peak value at \mbox{$\sim$16 T} and the center of the plateau-like region at \mbox{$\sim$23 T} is consistent with the factor $(\nicefrac{7}{9})^2/(\nicefrac{1}{2})^2$ expected for a transition between the $\nicefrac{2}{9}$ and $\nicefrac{1}{2}$ plateaus.
Finally, the magnetic (100) reflection disappears above 30 T when the field-polarized state is reached. Due to the (200) and (110) reflections having a sizeable nuclear contribution and being only and mostly, respectively, sensitive to the ferromagnetic $c$ component, the magnetic transitions are less pronounced in the intensity evolution, especially below 20 T where the antiferromagnetic in-plane component dominates. \newline

\begin{figure}
	\includegraphics[width=0.49\textwidth]{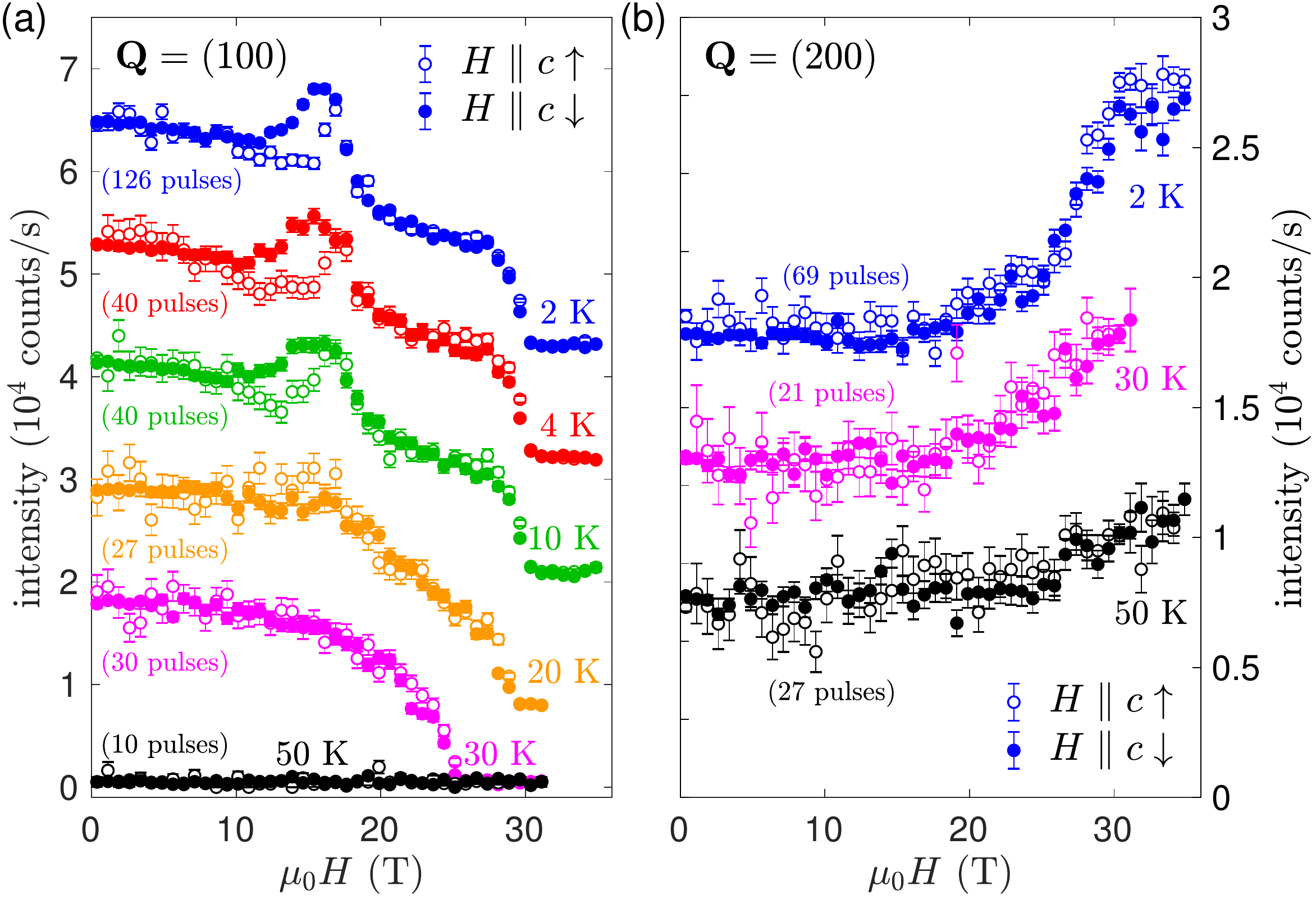}
	\caption{\label{fig:fielddep2} Magnetic field dependence of neutron diffracted intensities at: (a) $\mathbf{Q}$ = (1 0 0) and (b) $\mathbf{Q}$ = (2 0 0) in fields up to 35 T for different temperatures between 2 and 50 K. The magnetic field was applied along the $c$ axis. Open and full symbols correspond to rising and falling fields, respectively. Data below 30 K are shifted on the vertical axis for clarity.}
\end{figure}

The observed counts of the peak maxima within the different phases were extracted from the field dependence (Fig.~\ref{fig:fielddep})  at field values 0 T ($M/M_S=0$), 15.7 T ($M/M_S=\nicefrac{2}{9})$, 17.6 T ($M/M_S=\nicefrac{1}{3}$), 22.9 T ($M/M_S=\nicefrac{1}{2}$) and 31 T ($M/M_S=1$), and converted to integrated intensities by multiplying by $FWHM\cdot \sqrt{2\pi}/\sqrt{8\cdot\ln(2)}$, where $FWHM$ is the full width at half maximum of a Gaussian fit to the respective peak profile at $T$ = 2 K. The counts were furthermore corrected for the Lorentz factor and for the background by subtracting the (100) intensity at the highest applied field, where this particular Bragg reflection is extinct. The calculated integrated intensities were then converted back to peak amplitudes for direct comparison in Fig.~\ref{fig:fielddep}. \newline
The temperature dependence of the observed anomalies between 2 and 50 K are shown in Fig.~\ref{fig:fielddep2} and are consistent with the magnetic phase diagram derived from magnetization measurements and reported in Ref.~\cite{yos2008} and Fig.~\ref{fig:pd}.

\subsection{High-field magnetic structures}
\label{sec:modelling}
\begin{figure*}
	\includegraphics[width=0.99\textwidth]{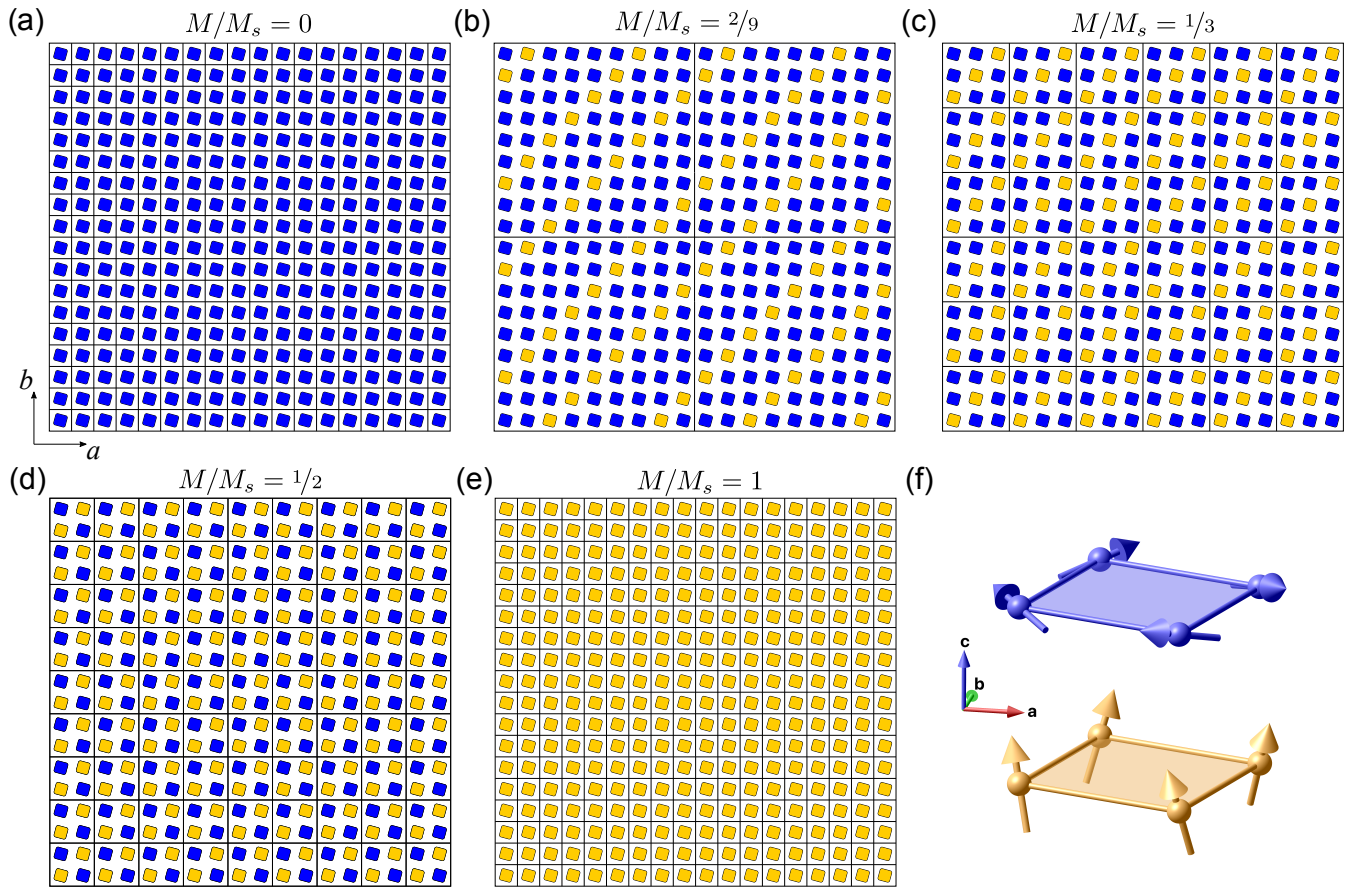}
	\caption{\label{fig:models} Magnetic structure models for the different phases in TbB$_4$, where a blue (orange) square represents a 4-spin-plaquette with magnetic moments within (perpendicular to) the $a$-$b$ plane. Black lines denote the magnetic unit cells. (a) zero-field magnetic structure with all magnetic moments lying within the $a$-$b$ plane, (b) $\nicefrac{2}{9}$-magnetization plateau phase with a 9$\times$9 supercell, in which polarized spin-plaquette stripes are alternately separated by 3 or 4 non-polarized stripes (note that this separation cannot be deduced from the data at hand and that this illustration would correspond to repelling polarized stripes), (c) $\nicefrac{1}{3}$-magnetization phase with a 3$\times$3 supercell featuring polarized plaquettes along the diagonal of each magnetic unit cell, (d) $\nicefrac{1}{2}$-magnetization phase with a 2$\times$2 supercell, (e) fully-polarized state with magnetic moments along the $c$ axis. Note that in (b) - (d) the magnetic moments represented by dark (light) squares do not lie fully within (perpendicular to) the $a$-$b$ plane and that they produce superstructure reflections at $\mathbf{Q}$ = ($\nicefrac{1}{9} $ $\nicefrac{1}{9}$ 0), $\mathbf{Q}$ = ($\nicefrac{1}{3} $ $\nicefrac{1}{3}$ 0) and $\mathbf{Q}$ = ($\nicefrac{1}{2}$ $\nicefrac{1}{2}$ 0), respectively, expressed with the conventional 1$\times$1 unit cell. (f) Detailed view on two 4-spin-plaquettes used as building blocks in the models (a)-(e). The upper (blue) one consists of magnetic moments predominantly aligned in the $a$-$b$ plane with a non-zero ferromagnetic $c$ component for the intermediate structures, while the lower (orange) one is almost fully polarized in (b)-(d) and fully polarized in (e).}
\end{figure*}

Here, we propose the $XY$-Ising spin mixture model which was employed to describe the half-magnetization plateau and can be extended to the $\nicefrac{2}{9}$ and $\nicefrac{1}{3}$-magnetization plateaus.
We have used the zero-field magnetic structure derived from the four-circle diffraction experiment as a starting point for the analysis of the field data. All refined structural and magnetic parameters from Sec.~\ref{sec:zerofield} were kept fixed and only the scale and extinction parameters were adapted to the IN22 experiment using the observed intensities at $H$ = 0 T. In the following, this adapted scale factor was fixed and the moment size was constrained to the zero-field value (note that the scale factor was correctly converted when working in magnetic supercells, e.g. it was divided by 16 for a 2$\times$2 cell). For the $\nicefrac{1}{2}$-magnetization state we have used the proposed mixture of $XY$- and Ising-type models \cite{yos2009} which we have further adapted to the $M/M_S=\nicefrac{2}{9}$ and $M/M_S=\nicefrac{1}{3}$ states by using a 9$\times$9 cell with 18 polarized spin plaquettes out of 81 and a 3$\times$3 cell with 3 polarized spin plaquettes out of 9, respectively. In all plateau phases only the $c$-component of one spin belonging to a polarized square plaquette was refined. The remaining spins were constrained to that value as well as the spins of the non-polarized plaquettes by defining a fixed magnetization. As mentioned above, the clear increase of the (100) reflection when entering the first plateau can only be explained by an increasing $b$ spin component due to the form of the magnetic structure factor. Because of the limited data set of 3 reflections the spins were fixed along the diagonal for the plateaus phases. It is a reasonable assumption that the nuclear structure may be affected by the applied field due to magnetostriction effects and the first-order transition between magnetic structures of different modulations most probably leads to a redistribution of magnetic domains which has an influence on the extinction effects. To cover these probably complex changes, a single parameter - the extinction coefficient $x_{11}$ (with the constraint $x_{11} = x_{22}$ due to the 90$^\circ$ twins) - was refined giving a maximum number of 2 parameters. For the fully polarized state the total spin of 8.94~$\mu\ut{B}$ was set along the $c$ axis and only the extinction parameter was refined. The results are shown as orange circles in Fig.~\ref{fig:fielddep} revealing a remarkably good agreement with the observed field dependence taken into account that all nuclear and magnetic structure changes were addressed by only one parameter each. Our proposed 9$\times$9 and 3$\times$3 supercell models for the $\nicefrac{2}{9}$- and $\nicefrac{1}{3}$-magnetization plateaus are in good agreement with the observed data. The refined parameters are shown in Tab.~\ref{tab:magparams} and the 5 magnetic structure models are illustrated in Fig.~\ref{fig:models} focusing on the distribution of (non)polarized spin plaquettes and different supercells.
\begin{table}
	\caption{Parameters of the magnetic structure analysis giving the best agreement for the different phases in \ch{TbB4}, which are expressed by their magnetization ratio $M/M_S$. The size of the magnetic moments is fixed to that obtained from the zero-field experiment in four-circle geometry. $\mu_1$ denotes a magnetic moment in a polarized plaquette, whereas $\mu_2$ refers to a spin plaquette which is predominantly oriented in-plane. }
	\label{tab:magparams}
	\begin{ruledtabular}
		\begin{tabular}{ccccc}
			$M/M_S$  &  $\mu_1\parallel c$ ($\mu\ut{B})$ &  $\mu_2\parallel c$ ($\mu\ut{B})$   &  $x_{11}=x_{22}$ & $R\ut{F}$   \\ \hline
			0  &  -  &   0 &  0.15(3) & 4.59 \\
			$\nicefrac{2}{9}$  &  8.4(3)  &   0.1(3)  &  0.09(2) & 5.37 \\
			$\nicefrac{1}{3}$  &  8.2(2)  &   0.4(2)  &  0.15(7) & 4.72 \\
			$\nicefrac{1}{2}$ &  8.4(2)  &   0.5(2)  &  0.4(1) & 6.16 \\
			1  &  8.94  &   - &  0.54(1) & 0.84 \\
		\end{tabular}
	\end{ruledtabular}
\end{table}

\begin{figure*}
	\includegraphics[width=0.99\textwidth]{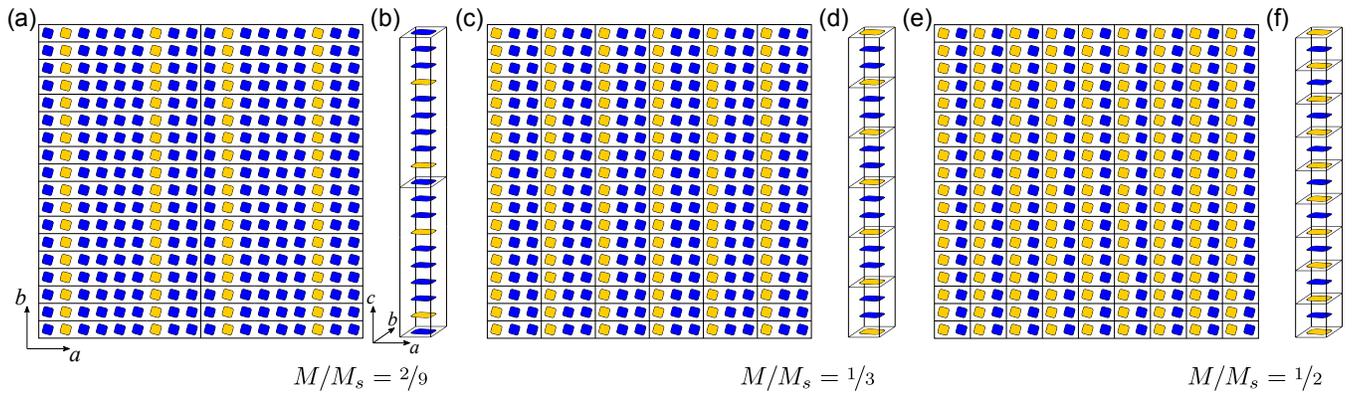}
	\caption{\label{fig:altmodels} Alternative models which explain the observed peak intensities of the plateau phases equally well as the models shown in Fig.~\ref{fig:models}(b) - (d). (a) and (b) are 9$\times$1$\times$1 and 1$\times$1$\times$9 supercells, respectively, and yield exactly the same structure factors for the 3 reflections which were followed as a function of magnetic field (Fig.~\ref{fig:fielddep}).  (c) and (d) are 3$\times$1$\times$1 and 1$\times$1$\times$3 supercells, respectively, and constitute alternative models for the $\nicefrac{1}{3}$ magnetization plateau phase. (e) and (f) are possible models for the $\nicefrac{1}{2}$-magnetization plateau and consist of 1$\times$2$\times$1 and 1$\times$1$\times$2 supercells, respectively. Note that the in-plane modulation can exist along the $b$ axis as well due to the presence of 90$^\circ$ twins.}
\end{figure*}

It can be seen that the plateau phases are characterized by diagonal stripes consisting of spins polarized along the $c$ axis by the applied magnetic field. While only regular arrangements of polarized stripes and non-polarized spacers are possible for the 2$\times$2 and 3$\times$3 supercells [Fig.~\ref{fig:models}(c) and (d)], several scenarios exist for the 9$\times$9 supercell, in which different spacings between polarized stripes are conceivable (all being undistinguishable by neutron scattering).\newline\newline
Alternative models exist which explain the observed peak intensities in the plateaus phases equally well without having additional information of further Bragg reflections as will be discussed in Sec.~\ref{sec:conclusion}. What basically distinguishes these models is the type of modulation of the superstructure which is summarized in Fig.~\ref{fig:altmodels}. While superstructures with multiple cells along the $a$ and $b$ axes were used in the previously proposed analysis, which yield patterns of diagonal stripes of polarized spin-plaquettes (Fig.~\ref{fig:models}), it is also conceivable that stripes are aligned along the $a$ axis or equivalently along the $b$ axis due to the presence of 90$^\circ$ twins. The building block of the stripe models is also debatable as polarized spin dimers, similar to TmB$_4$ \cite{sie2008} instead of square plaquettes proposed here, lead to similar patterns. Another possibility is to stack entire planes of polarized and unpolarized spin plaquettes leading to magnetic unit cells which have the same in-plane periodicity as the nuclear unit cells but with a multiple of the $c$ axis. All these models yield exactly the same intensities for the 3 Bragg reflections investigated here. However, stacked ferromagnetic planes ($\mu\parallel c$) in an up-up-down fashion as proposed for HoB$_4$ \cite{bru2017} can definitely be excluded since such a model would produce zero magnetic intensity for the (100) reflection at intermediate field values.

\section{Discussion}
\label{sec:conclusion}
Our pulsed-field neutron diffraction experiments with improved time resolution and precise control of the sample temperature (in comparison to Ref.~\onlinecite{yos2009}) offer new microscopic insight into the  magnetization plateaus behavior of the Shastry-Sutherland system TbB$_4$. While we can confirm the zero-field, half- and full-magnetization plateau structures, the magnetic structures of the $\nicefrac{2}{9}$- and the $\nicefrac{1}{3}$-magnetization plateaus are reported for the first time which can be explained - in analogy to the $M/M_S=\nicefrac{1}{2}$ plateau - by a 9$\times$9 and a 3$\times$3 magnetic supercell, respectively. However, the scattered neutron intensity as a function of applied field does not provide a microscopic proof for the appearance of the $\nicefrac{4}{9}$- and the $\nicefrac{7}{9}$-plateaus, which - according to the macroscopic magnetization data - should be established at field values of approximately 19 T and 27 T, respectively. This may be related to the transition dynamics, the phase stability and the short duration of the magnetic field pulse. \newline

In the plateau phases the \textit{excess} magnetization is carried by the system via stripes of polarized spin plaquettes, where the stripe density increases with increasing magnetization. This phenomenon is reminiscent of cuprates,  manganites, nickelates and cobaltates in which hole or electron doping forms checkerboard or stripe patterns. No conclusion can be drawn at this point concerning the stripe mobility or interaction in TbB$_4$ without further experimental and theoretical effort. While the stripe order necessarily reveals a rather simple pattern with regular spacings for the $\nicefrac{1}{2}$- and $\nicefrac{1}{3}$-plateaus, the questions of how they are distributed e.g. in the $\nicefrac{2}{9}$-plateau arises. Depending on the type of interaction between the polarized stripes it can be energetically more favorable to separate them as far as possible, which is shown in Fig.~\ref{fig:models}(b), on the other hand, attracting stripes would reveal a pattern of 2 polarized stripes separated by 7 non-polarized ones. Note that both possibilities yield the same intensities for integer reflections (referring to the conventional unit cell) and therefore can only be distinguished based on their superstructure reflection patterns.  \newline
We have identified additional, differently modulated superstructures equally reproducing the observed peak intensities of integer ($hkl$) reflections, which therefore cannot be distinguished based on the data at hand. While the in-plane modulations only differ concerning the stripe pattern of polarized plaquettes (along the diagonal vs. along the $a$ or $b$ direction), the out-of-plane modulation implies an alternate stacking of differently polarized planes which would reveal a completely distinct coupling scheme between plaquettes in the basal plane as well as perpendicular to it. The models shown in Fig.~\ref{fig:models}(b) - (d) should produce purely magnetic superstructure reflections, from which we have simulated the fundamental \mbox{$\mathbf{Q}$ = ($\delta$ $\delta$ 0)} reflections to be the strongest ones for the respective superstructures with \mbox{$\delta$ = $\nicefrac{1}{9}$}, $\nicefrac{1}{3}$ or $\nicefrac{1}{2}$, with the $(\nicefrac{1}{2}\ \nicefrac{1}{2}\ 0)$ reflection being approximately 60\% of the (100) intensity and therefore easily detectable with the same experimental setup. On the other hand, models shown in Fig.~\ref{fig:altmodels}(a), (c) and (e) would show purely magnetic scattering of comparable strength at ($\delta$ 0 0) or equivalently at (0 $\delta$ 0) positions, while Fig.~\ref{fig:altmodels}(b), (d) and (e) yield $\mathbf{Q}$ = (0 0 $\delta$) superstructure reflections. Magnetic superstructure reflections at ($\nicefrac{1}{8}$ 0 0) and (0 $\nicefrac{1}{8}$ 0) positions were indeed found in the first plateau phase of TmB$_4$ (Ref.~\onlinecite{sie2008}), but this system is fundamentally different from TbB$_4$ due to its strong Ising character even at zero magnetic field and spin-dimer properties in the field-induced phases, which, nevertheless, were also described by stripe patterns. It is therefore crucial to further explore the magnetic scattering in reciprocal space to discern the different proposed models and understand the complex magnetization behavior in TbB$_4$.

\begin{acknowledgments}
	The CEA-CRG Grenoble and the ILL are greatly acknowledged for granting the beam time for these experiments.
	The authors are very grateful to X. Tonon and E. Leli\`{e}vre-Berna for their active support with the cryogenics. The authors would like to thank H. Nojiri for fruitful discussions and O. Fabelo for complementary single crystal x-ray measurements.
	This work was financially supported by the French National Research Agency (ANR project MAGFINS: Grant N$^\circ$. ANR-10-BLN-0431) and by
	the program Investissements d'Avenir ANR-11-IDEX-0002-02 (reference ANR-10-LABX-0037-NEXT).
\end{acknowledgments}

\appendix
\bibliographystyle{apsrev4-2}

\end{document}